\documentclass{ws-procs961x669}
\usepackage{graphicx}

\begin{document}

\wstoc{For proceedings contributors: Using World Scientific's\\ WS-procs961x669 document class in \LaTeX2e}
{F. Camilloni}
\index{author}{Camilloni, F.}

\title{
Blandford-Znajek power as a strong-gravity signature }
\author{F. Camilloni}
\address{
Institut f\"ur Theoretische Physik, Goethe Universit\"at,
\\
Frankfurt am Main, Germany, Max-von-Laue-Str. 1, 60438
\\
E-mail: camilloni@itp.uni-frankfurt.com}

\begin{abstract}
The Blandford-Znajek mechanism is an electromagnetic manifestation of the Penrose process that currently constitutes the best theoretical candidate to explain the launching of relativistic jets by black holes.
In this talk we offer a modern review about the Blandford-Znajek mechanism and the analytic construction of black hole magnetospheres. Higher order perturbative corrections are crucial in order to produce results that are complementary to numerical simulations when the black hole is in the high-spin regime, and can potentially predict new features about the non-perturbative structure of the Blandford-Znajek theory. Moreover, we show by means of an explicit example that these perturbative corrections depend in a non-degenerate manner on the underlying theory of gravity considered, enabling one to use the BZ power emitted as a strong-gravity signature to test General Relativity against alternative theories of gravity on future horizon-scale observations.
\end{abstract}


\bodymatter

\section{Introduction}\label{sec: intro}

The mechanism responsible for the emission of jets by supermassive Black Holes (BHs) is one of the outstanding questions in the field of relativistic astrophysics.
New observational perspectives were recently opened by the Event Horizon Telescope (EHT), with the potential to shed a new light on the complex interplay of matter and electromagnetic fields subject to the extreme conditions characterizing the BH environment.
In fact the observations already revealed some of the features expected to be crucial in the physics of jet emission: the polarized images of M87* produced by EHT confirmed that strong and ordered magnetic fields are present at the horizon scale~\cite{EventHorizonTelescope:2019ds,EventHorizonTelescope:2019pgp,EventHorizonTelescope:2021srq,EventHorizonTelescope:2022wkp}, whereas comparisons with synthetic templates obtained by means of GRMHD simulations fostered models in which the BH is rapidly spinning and the jet is launched via the Blandford-Znajek (BZ) mechanism~\cite{Blandford:1977ds}.

Since its first theoretical formulation, and well before the strong indications provided by EHT, the BZ mechanism has always been considered a favourite theoretical candidate to explain how relativistic jets are launched and sustained by BHs immersed in strongly magnetised environments, and many research efforts have been devoted to clarify its dynamics and its relevance in BH physics~\cite{10.1046/j.1365-8711.2001.04863.x,Uzdensky:2003cy,Uzdensky:2004qu,Komissarov:2004ms,Tanabe:2008wm,Tchekhovskoy:2011zx,Nathanail:2014aua,Gralla:2015vta,Camilloni:2020qah,Camilloni:2020hns,Armas:2020mio,Camilloni:2022kmx}.

According to our modern understanding, the BZ mechanism is an electromagnetic manifestation of a generalised Penrose process~\cite{1971NPhS..229..177P,Lasota:2013kia}, in which the active component operating the energy and angular momentum extraction is a strong magnetosphere that spins together with the BH. The nature of the BZ mechanism is thus deeply rooted in general relativity, and a natural question to ask in this regard is whether strong-gravity signatures can be derived by analyzing the power emitted in this process.
In order to address this question it is therefore crucial to clarify which elements contribute to the expression that characterize the BZ power.

Intuitively, due to its Penrose process-like nature, it is to be expected that the faster the BH spins the more energy will be available to be extracted. Moreover, since the energy extraction is operated by the magnetic environment in which the BH is immersed, one also expects that the greater is the number of open magnetic field lines penetrating the horizon the more energy can be transfered from the BH to the infinity. A leading order approximation for the BZ power, already derived in the seminal BZ paper~\cite{Blandford:1977ds}, is indeed considered to be $\dot{E}_+\simeq\kappa\left(2\pi \Psi_H\right)^2\Omega_H^2$, where $\kappa$ is a numerical coefficient associated to the magnetospheric topology, $(2\pi\Psi_H)$ represents the flux of open magnetic field lines threading the BH horizon, and $\Omega_H$ is the BH angular velocity~\cite{2010ApJ...711...50T,Tchekhovskoy:2011zx}.
The scaling $\dot{E}_+\sim\Omega_H^2$, in particular, is typically regarded as a distinguishing signature of the BZ mechanism, often employed in numerical investigation to identify the process.
Nonetheless, analytical~\cite{Camilloni:2022kmx} as well as numerical~\cite{2010ApJ...711...50T} studies agree that for BHs in the high spinning regime the contribution of an additional \emph{high-spin factor} $f(\Omega_H)$ becomes relevant. A more accurate expression for the BZ power extracted that accounts for its presence, thus, reads
\begin{equation}
\label{eq: BZE}
	\dot{E}_+=\kappa\left(2\pi \Psi_H\right)^2\Omega_H^2~f(\Omega_H)~~.
\end{equation}
The function $f(\Omega_H)$ can be computed either numerically, by fitting data extracted from the simulations~\cite{2010ApJ...711...50T}, or analytically by combining perturbative approaches with matched asymptotic expansion schemes~\cite{Camilloni:2022kmx}. The latter method has been employed to derive an explicit sixth order expression for a split-monopole magnetospheres in the Kerr metric, that has been shown to be featured with logarithmic terms
\begin{equation}
\label{eq: f}
\begin{split}
	f(\Omega_H)\simeq 1 + 1.38 (M\Omega_H)^2 &-11.25(M\Omega_H)^4 +1.54|M\Omega_H|^5 
	\\
	&+\Big[11.64 -0.17 \log|M\Omega_H|\Big](M\Omega_H)^6 +\dots ~~,
\end{split}
\end{equation}
and well agrees with simulations even for BHs spinning closely to the Thorne limit~\cite{Camilloni:2022kmx}.
From Eq.~\eqref{eq: f} one can immediately observe that if the BH is slowly spinning the usual quadratic scaling in $\Omega_H$ is recovered as the leading order term of the series.
Moreover, numerical indications give support to the fact that $f(\Omega_H)$ only weakly dependends on the details of the magnetospheric configuration~\cite{2010ApJ...711...50T}, and thus one can expect the expression above to be characteristic of the BZ mechanism in the Kerr background.

Given our current understanding of the BZ mechanism it is natural to ask where the strong-gravity nature of this process manifests at the level of Eq.~\eqref{eq: BZE}, and whether such an expression enables us to test the Kerr paradigm in future horizon-scale observations at EHT, with the aim of understanding if the BHs emitting relativistic jets are well described by the Kerr metric or not.\\

The main goal here~\cite{Camilloni:2023wyn} is to motivate and provide theoretical support to the idea that the high-spin factor $f(\Omega_H)$ is not only necessary to correctly model the energy extraction for rapidly spinning BHs, but it is also the quantity where the details of the background metric enter unambiguously in the power~\eqref{eq: BZE}. In this regards we notice that truncating the BZ power at the leading order, $\dot{E}_+\sim\Omega_H^2$, would not be useful to test the nature of a BH, since the angular velocity $\Omega_H$ is typically a quantity degenerate in the parameters that characterize the spacetime metric~\cite{Dong:2021yss,Camilloni:2023wyn}.

We support this hypothesis by exploring analytically a concrete example in which the BZ mechanism operates in the context of an alternative theory of gravity, and by studying how the power extracted deviates from the expression derived in the Kerr metric~\cite{Camilloni:2023wyn}.

As a prototype of a non-Kerr background we consider a stationary and axisymmetric BH solution of Scalar-Tensor-Vector Gravity (STVG)~\cite{Moffat:2005si} -- also known in the literature as MOdified Gravity (MOG) -- called Kerr-MOG~\cite{Moffat:2014aja}. STVG constitutes a covariant extension of GR still not discharded by observations, in which an additional repulsive Yukawa potential alters the asymptotic value of the effective gravitational coupling constant compared to its value close to the source. In this way the theory provides a possible justification to phenomena like the galaxy rotation curves without invoking dark matter~\cite{Brownstein:2005zz}. 
We refer the reader to the literature for extensive details about STVG, while here we limit to stress that an additional deformation parameter $q=(G_\infty-G_N)/G_N$ enters the Kerr-MOG solution, accounting for deviation from the Newton constant at large scales~\cite{Moffat:2014aja}.
Explicitly, the Kerr-MOG metric in Boyer-Lindquist coordinates is given by  
\begin{equation}
ds^2=-\frac{\Delta_q \Sigma}{\Pi}dt^2+\frac{\Pi\sin^2\theta}{\Sigma}\left(d\phi-\omega_q dt\right)^2+\frac{\Sigma}{\Delta_q}\,dr^2 +\Sigma \,d\theta^2~~,
\end{equation}
with $\Sigma=r^2+a^2\cos^2\theta$, $\Pi=(r^2+a^2)^2-a^2\Delta_q \sin^2\theta$, $\Delta_q=r^2-2M_{q}  r+a^2+ q/(1+q) M_q^2$ and $\omega_q =a(2M_qr-q/(1+q)^2M_q)/\Pi$. The Arnoitt-Deser-Misner mass and angular momentum depend on the deformation parameter $q$ according to $M_{q}=(1+q)M$ and $J=a M_q$~\cite{Sheoran:2017dwb}. The coordinate singularities are located at $r_\pm=M_q\left(1\pm\sqrt{1/(1+q)-(a/M_q)^2}\right)$ and the BH angular velocity is $\Omega_H(q)=a/M_q\left(2r_+-q/(1+q)M_q\right)^{-1}$. From here we see that the angular velocity is degenerate with respect to the spin $a$ and the deformation parameter $q$, and therefore the quadratic scaling that characterize the leading order expression of the BZ power, $\dot{E}\sim\Omega_H^2$, is not useful to perform tests on the Kerr paradigm.
\\

Having clarified the motivations and the context in which this research proceeds, in the following sections we offer a short review of the BH magnetospheric problem and of the analytic techniques that can be employed to construct consistent models. These approaches can be systematically adapted in non-Kerr backgrounds to derive expression of the BZ power emitted in alternative theories of gravity, as we show in the last section by analyzing the case of a Kerr-MOG BH.

\section{The black hole magnetospheric problem}\label{sec: magn }

We define a magnetosphere as a region in the BH environment in which the plasma only plays a passive role in screening longitudinal electric fields and in sustaining the magnetic fields, that instead dominate the dynamics in the curved BH spacetime geometry~\cite{Gralla:2014yja}. This condition is captured by assuming that the electromagnetic fields satisfy the force-free condition $T^{\mu\nu}_{\rm em}\gg T^{\mu\nu}_{\rm mat}$, implying that the equations of Force-Free Electrodynamics (FFE) are covariantly represented by the following system of non-linear differential equations in a curved background metric~\cite{PhysRevE.56.2181,PhysRevE.56.2198} 
\begin{equation}
\label{eq: FFE}
	{F^\mu}_\nu\nabla_\rho F^{\rho\nu}=0~~,~~\nabla_\mu \star F^{\mu\nu}=0~~.
\end{equation}
Since the plasma is present but dynamically secundary compared to the electromagnetic fields, henceforth we identify a magnetosphere with the field strength $F_{\mu\nu}$.

If one assumes that the magnetosphere respects the symmetry of a stationary and axisymmetric background metric, it is possible to express the field strength in terms of the magnetic flux $\Psi(r,\theta)$, that is constant along the poloidal projection of the magnetic field lines, the poloidal current $I(\Psi)$, related to a toroidal magnetic field, and the angular velocity of the magnetic field lines $\Omega(\Psi)$, that is responsible for transverse electric fields. The structure of a stationary and axisymmetric magnetosphere in the Kerr-MOG background can be derived by adapting the standard expression adopted in the Kerr metric, and is represented by a $2$-form of the kind~\cite{Gralla:2014yja}
\begin{equation}
	F=d\Psi\wedge (d\phi-\Omega(\Psi) dt)-I(\Psi)~\frac{\Sigma}{\Delta_q\sin\theta}\,dr\wedge d\theta~~.
\end{equation}
It is easy to observe that the field above automatically satisfies the Bianchi identity, and that the equations of FFE \eqref{eq: FFE} reduce to a single BH Grad-Shafranov equation, that in the Kerr-MOG metric reads~\cite{Camilloni:2022kmx}
\begin{equation}
	\label{eq: GS_eq}
    \eta_\mu \partial_r \Big( \eta^\mu  \Delta_q \sin \theta \,  \partial_r \Psi \Big) + \eta_\mu \partial_\theta \Big( \eta^\mu  \sin \theta \,  \partial_\theta \Psi \Big) + \frac{\Sigma}{\Delta_q\sin \theta} I \frac{d I}{d\Psi} =0~~,
\end{equation}
with $\eta_\mu=(d\phi-\Omega(\Psi) dt)_\mu$.
The equation above is in general a second-order quasilinear Partial Differential Equation (PDE) for $\Psi(r,\theta)$, but becomes first order at the horizon $r\to r_+$, at the infinity $r\to\infty$, and at two distinct light surfaces, solutions of $\eta^\mu\eta_\mu=0$. In order to guarantee that the solution is regular at these four critical surfaces, the BH Grad-Shafranov equation must be supplemented with a set of regularity conditions. At the horizon and at infinity the \emph{Znajek conditions} enforce the regularity of the fields measured by a freely-falling observer, in the Kerr-MOG background
\begin{equation}
    \label{eq: ZC}
    I_+(\theta)=\frac{2M_q r_+}{\Sigma_+}\sin\theta\left(\Omega_H-\Omega_+\right)\partial_\theta\Psi_+~~,~~
    I^\infty(\theta)=\sin\theta\,\Omega^\infty(\partial_\theta\Psi)^\infty~~.
\end{equation}
The following \emph{reduced stream equation}, instead, implies the smoothness of $\Psi(r,\theta)$ at the Inner and Outer Light Surface (ILS/OLS)  \cite{Uzdensky:2003cy,Uzdensky:2004qu,Gralla:2014yja,Camilloni:2022kmx} 
\begin{equation}
    \label{eq: stream_LS}
    \Delta_q\left(\eta_\mu \partial_r \eta^\mu\right)  \partial_r \Psi +\left( \eta_\mu \partial_\theta  \eta^\mu \right) \partial_\theta \Psi  + \frac{\Sigma}{\Delta_q \sin^2 \theta} I \frac{dI}{d\Psi} =0~~.
\end{equation}
The set of Eqs.~\eqref{eq: GS_eq},~\eqref{eq: ZC} and \eqref{eq: stream_LS} constitutes the complete system of  differential equations needed to close the BH magnetospheric problem, and allows to determine functional expressions for the field variables $\Psi(r,\theta)$, $I(\Psi)$ and $\Omega(\Psi)$.
\\

Once a specific magnetospheric configuration is known, the power extracted from the BH via the BZ mechanism can be computed by integrating over a sphere of fixed radius $r=r_c$ according to $\dot{E}(r_c)=2\pi\int_0^\pi I(r_c,\theta)\Omega(r_c,\theta)\partial_\theta\Psi(r_c,\theta)d\theta$~\cite{Gralla:2014yja}.
It is costumary to choose the horizon itself $r_c=r_+$ as surface to compute the energy extraction rate, so that it is possible to use the Znajek condition \eqref{eq: ZC} to write
\begin{equation}
	\dot{E}_+=2\pi\int_0^\pi \frac{2M_q r_+}{\Sigma_+}\sin\theta \Omega(r_+,\theta) (\Omega_H-\Omega(r_+,\theta))\partial_\theta\Psi(r_+,\theta) d\theta~~,
\end{equation} 
from which one can see that the energy is extracted ($\dot E_+>0$) if the magnetosphere spins together with the BH, but with a rate smaller than the BH angular velocity $\Omega_H>\Omega_+>0$.

\section{Analytic approaches for spinning black hole magnetospheres}\label{sec: pert }
Given the non-linear structure of the Grad-Shafranov equation \eqref{eq: GS_eq}, the system characterizing the BH magnetospheric problem can be approached analytically by means of perturbative techniques.
In their seminal paper Blandford and Znajek~\cite{Blandford:1977ds} considered the dimensionless BH spin, $\chi=a/M_q$ in the Kerr-MOG background, as a natural parameter to develop a perturbation theory for $\chi\ll1$.

The BZ perturbative approach \cite{Blandford:1977ds,Armas:2020mio,Camilloni:2022kmx} regards a vacuum field on a static BH background as a seed solution for a magnetostatic field akin to $F_{\mu\nu}\sim \mathcal{O}(\chi^0)$. Perturbative force-free corrections follow by assuming a plasma current emerges if the BH is spinning, so that $j^\mu=\nabla_\nu F^{\mu\nu}\sim\mathcal{O}(\chi)$.
This is compatible with field variables expanded according to
\begin{equation}
\label{eq: expansion}
\begin{split}
	&\Psi(r,\theta)=\psi_0+\chi^2~\psi_2+\mathcal{O}(\chi^4)~~,
	\\
	I(r,\theta)=\chi~ &\frac{i_1}{M_q} +\mathcal{O}(\chi^3) ~~,~~ \Omega(r,\theta)=\chi~ \frac{\omega_1}{M_q}+\mathcal{O}(\chi^3)~~,
\end{split}
\end{equation}
with $\psi_0$ representing a vacuum poloidal magnetic field in the static background. Intuitively we can think that when $\chi\neq0$ the magnetosphere is dragged into the spinning motion by the BH, so that also the magnetic field lines rotate, $\Omega\sim \mathcal{O}(\chi)$, and additional toroidal components are acquired, $I\sim \mathcal{O}(\chi)$.

The vacuum seed solution follows upon considering the leading order term of Eq.~\eqref{eq: GS_eq}, expanded via Eq.~\eqref{eq: expansion}. This is a homogeneous PDE akin to $\hat{\mathcal{L}}\psi_0=\mathcal{O}(\chi)$, with the operator in the static MOG background being defined as~\cite{Camilloni:2023wyn}
\begin{equation}
\label{eq: L}
    \hat{\mathcal{L}}=\frac{1}{\sin\theta}\partial_r\left[\left(1-\frac{2M_q}{r}+\frac{q}{(1+q)}\frac{M_q^2}{r^2}\right)\partial_r\right]+\frac{1}{r^2}\partial_\theta\left(\frac{1}{\sin\theta}~\partial_\theta\right)~~.
\end{equation}
The operator can be separated and a generic solution is determined by superposing radial and angular harmonics, $\psi_0=c_0+\sum_\ell (c_u~U_\ell(r)+c_v~V_\ell(r)) \Theta_\ell(\theta)$~\cite{Armas:2020mio}. 
It is interesting for our purpose to keep track of the deformation parameter $q$ which, as it is clear from Eq.~\eqref{eq: L}, only affects the radial part of the operator. This means that the angular harmonics remain the same as in the standard Schwarzschild case, whereas the functional expression of the radial harmonics can be obtained in the Schwarzschild-MOG background upon solving a Heun type differential equation~\cite{Ronveaux:1995:HDE,2007MaCom..76..811M}. Two sets of radial eigenfunctions can be derived, explicitly given by~\cite{Camilloni:wyn}
\begin{align}
\nonumber
    U_\ell(w_q;w)&=\frac{(2M_q)^{\ell+1}}{\ell(\ell+1)}\frac{\Gamma(\ell+2)^2}{\Gamma(2\ell+1)}(1-w_q)^\ell H_l\left(\frac{1}{1-w_q},\frac{\ell(\ell+1)}{w_q-1},-(1+\ell),\ell,1,1;\frac{w-1}{w_q-1}\right)~~,
    \\ \label{eq: U_V}
    V_\ell(w_q;w)&=2M_q(2\ell+1)U_\ell(w_q;w)\int_w^\infty\frac{t^2}{(t-1)(t-w_q)U^2_\ell(w_q;t)}dt~~,
\end{align}
where $H_l$ denotes the Heun polynomials and we redefined a dimensionless radial coordinate as $w=r/M_q(1+1/\sqrt{1+q})^{-1}$ and $w_q=(1+q-\sqrt{1+q})/(1+q+\sqrt{1+q})$.
For $q\to0$ one recovers the hypergeometric functions of the standard Schwarzschild vacuum fields~\cite{Gralla:2015vta}.
Eq.~\eqref{eq: U_V} provides a complete set of functions, with $U_\ell(w_q,w)$ regular at the horizon and divergent at infinity for $\ell>1$, and viceversa for $V_\ell(w_q,w)$.

Different vacuum seed solutions in the static background differs for their asymptotic boundary conditions~\cite{Grignani:2019dqc}. For instance, monopole fields $\psi_0(\theta)=1-\cos\theta$ can be obtained by demanding finiteness of the magnetic flux at the radial infinity.  Asymptotically vertical vacuum fields are instead given by the $\ell=1$ harmonic, that after suitable normalization leads to $\psi_0(r,\theta)=[(1+q)r^2-qM_q^2]/[2M_q^2(1+\sqrt{1+q})]\sin^2\theta$ in the Schwarzschild-MOG case.

Starting from a vacuum seed solution, higher order magnetospheric configurations can be constructed by perturbatively solving in $\chi$ the set of Eqs.~\eqref{eq: GS_eq},~\eqref{eq: ZC} and \eqref{eq: stream_LS} via the expansion \eqref{eq: expansion}.
At each order in $\chi$ the Grad-Shafranov equation is written as a non-homogeneous PDE, $\hat{\mathcal{L}
}\psi_n(r,\theta)=\mathcal{S}(r,\theta; \psi_{k<n},i_{k<n},\omega_{k<n})$, and the perturbative correction to the magnetic flux can be derived via variable separation and harmonics superposition $\psi_n(r,\theta)\sim R_n^{(\ell)}(r)\Theta_{\ell}(\theta)$, for instance by making use of the Green function method~\cite{Blandford:1977ds,Camilloni:2022kmx}.
\\

\begin{figure}[t]
	\centering
  \includegraphics[width=0.95\textwidth]{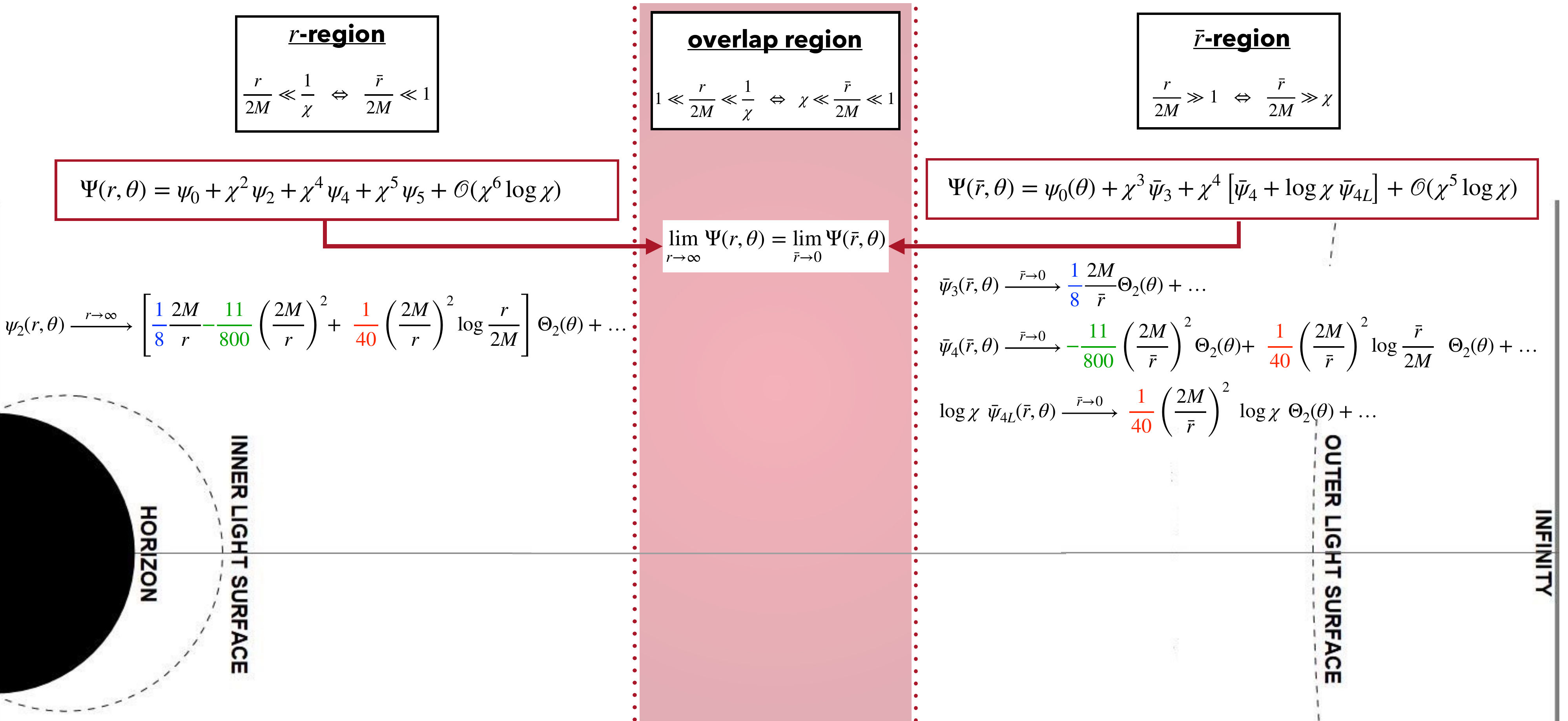}
  \caption{Matched asymptotic expansion scheme for a monopolar magnetosphere in the Kerr background~\cite{Camilloni:2022kmx}. In the picture we provide an example concerning the coefficient $\psi_2$, by highlighting with colours the matching between the asymptotic expansions in the $r$ and in the $\bar{r}$ regions.}
  \label{fig: MAE}
\end{figure}

Recently it was shown that the perturbation theory in general breaks down at the outer light surface~\cite{Armas:2020mio}, that scales in a non-perturbative manner as $r\sim\chi^{-1}$. In order to construct consistent higher-order magnetospheric solutions, it is therefore necessary to enhance the BZ perturbative approach by making use of matched asymptotic expansion schemes~\cite{Armas:2020mio,Camilloni:2022kmx}. This makes use of an alternative radial coordinate $\bar{r}=\chi r$, solely designed to keep the position of the outer light surface fixed while performing the expansion for $\chi\ll1$. Introducing a second radial coordinate effectively amounts to divide the space outside the horizon in an $r$-region, that encompass the horizon and the inner light surface, and in an $\bar{r}$-region, enclosing the outer light surface and the infinity. 
In each region it is possible to independently expand the field variables. In particular, by focusing on the magnetic flux one has that
\begin{equation}
	\begin{split}
		\Psi(r,\theta)&=\psi_0+\chi^2 \psi_2+\chi^4 \psi_4+\mathcal{O}(\chi^5)~~~~~~~~,~~~~~~~~~r/(2M)\ll \chi^{-1}~,~\bar{r}/(2M)\ll 1
		\\
		\Psi(\bar{r},\theta)&=\psi_0+\chi^3 \bar{\psi}_3+\chi^4 (\bar{\psi}_4+\bar{\psi}_{4L}\log\chi)+\mathcal{O}(\chi^5)~~,~~r/(2M)\gg 1~,~\bar{r}/2M\gg\chi
	\end{split}
\end{equation}
and the single coefficients must be derived by perturbatively solving the magnetospheric equations, with the constraint that the two expansion above match each other at every order in the perturbation theory in the overlap region, $1\ll r/2M\ll \chi^{-1}$ and $\chi \ll \bar{r}/2M \ll 1$, where both the radial coordinates are suitably defined. Fig.~\ref{fig: MAE} illustrates how the matched asymptotic expansion scheme works in the context of a split-monopole magnetosphere in the Kerr geometry~\cite{Camilloni:2022kmx}.

The matched asymptotic expansion scheme has proven to be a crucial technique for the analytic construction of higher-order magnetospheric models that can complement the results obtained via numerical simulations, and to perform detailed computations of the high-spin factor $f(\Omega_H)$, as given in Eq.~\eqref{eq: f} for the case of a Kerr BH~\cite{Camilloni:2022kmx}.

\section{Blandford-Znajek power in modified gravity}
As motivated above, the analytic approaches we described can be adapted in order to construct consistent magnetospheric models in alternative theories of gravity, for instance in the Kerr-MOG background~\cite{Camilloni:2023wyn}.

Having the magnetospheric solution explicitly it is then possible to perturbatively compute the energy extraction rate at the horizon as $\dot{E}_+=\chi^2~\dot{E}^{(2)}_+(\psi_0,i_1,\omega_1)+\chi^4~\dot{E}^{(4)}_+(\psi_{0,2},i_{1,3},\omega_{1,3})+\mathcal{O}(\chi^6)$. A better convergence is obtained by converting the series in terms of the BH angular velocity~\cite{2010ApJ...711...50T}, which for a monopole field in the Kerr-MOG background leads to $\dot{E}_+=2\pi/3~\Omega_H^2 f_q(\Omega_H)$. The high-spin factor associated to a Kerr-MOG BH and computed at the next-to-leading order reads~\cite{Camilloni:2023wyn}
\begin{equation}
f_q(\Omega_H)=1+\frac{4}{5}M_q^2\Omega_H^2 ~\frac{\left(1+\sqrt{1+q}\right)^2}{1+q}\Bigg[1-\frac{\left(1+\sqrt{1+q}\right)^2}{1+q}R_2^H(q)\Bigg]+\mathcal{O}(\Omega_H^4)~~,
\end{equation}
with the function
\begin{equation}
R_2^H(q)=\frac{(1+w_q)^2}{2(1-w_q)w_q}\left[\frac{w_q(3\pi^2-47+2w_q)}{18}-\frac{1}{2}+Li_2(w_q)-\frac{1-w_q^2}{2w_q}\log(1-w_q)\right]~~.
\end{equation}
Notice that at the leading order, $\dot{E}_+=2\pi/3~\Omega_H^2+\mathcal{O}(\Omega_H^4)$, the BZ power for a monopole field has the same formal expression one would derive in the Kerr metric.
We stress that the BH angular velocity is a degenerate quantity, as an infinite number of combinations for the spin and the deformation parameters can be used to produce the same value of $\Omega_H$. Hence, the quadratic scaling $\dot{E}_+\sim\Omega_H^2$, typical of the BZ power truncated at the leading-order, cannot be useful to distinguish a Kerr BH from a possible MOG counterpart.

It is instead immediate to observe that the MOG deformation parameter $q$ enters in a non-trivial manner at the next-to-leading order in the definition of the high-spin factor $f_q(\Omega_H)$. For sufficiently high values of the BH spin and deformation parameters it is possible to appreciate differences in the energy extracted via the BZ mechanism. If we set $q\approx 2.45$, that is inside the range of possible values of $q$ compatible with supermassive BH observations~\cite{Perez:2020ndx}, and $M_q \Omega_H\approx 0.4$, the fractional difference between the power extracted in the MOG case compared to the one extracted in general relativity is of the order $\approx 15\%$~\cite{Camilloni:2023wyn}.

The specific example we provided by analyzing the Kerr--MOG case, thus, suggests that the high-spin factor $f(\Omega_H)$ depends on the underlying theory of gravity where the extraction process is set to operate, and that can in principle be used in horizon-scale observations combined with independent measurements of the BH spin to probe the spacetime metric and test the Kerr paradigm.

\section{Conclusions and outlooks}

The main point we aim to stress with this work is that the quadratic scaling $\dot E_+\sim\Omega_H^2$, often regarded as a signature of the BZ mechanism, is in reality not always representative for this process. An analytic study of the BH magnetospheric problem reveals that this is just the leading order term in a perturbative series for $\Omega_H$, that fails in capturing the energy extraction rate for BHs in the high-spin regime, as well as in identifyng unambiguously the spacetime metric due to the degeneracy that affects $\Omega_H$.

It is instead the high-spin factor $f(\Omega_H)$ in the BZ power extracted at the horizon, Eq.~\eqref{eq: BZE}, that contains crucial insights about the strong-gravity nature of the mechanism and about the energy extraction for BHs in the high spinning regime.
\\
In both cases an accurate knowledge of the expression for $f(\Omega_H)$ is crucial in order to access results that are comparable with the numerical simulations and to distinguish general relativity from alternative theories of gravity. In this regard we showed that the analytical methods described above can be extremely useful to derive expressions for this quantity and have control of the magnetospheric solution in the entire parameter space.

We stress that the BZ mechanism is a purely electromagnetic extraction process operated by a force-free magnetosphere surrounding a BH, and accordingly in the consideration made above we never referred to the presence of other elements in the BH environment, like for istance accretion discs.
While it is known that Magnetically-Arrested Disc (MAD) models can enhance the BZ efficiency of energy extraction by means of a dramatic increase of the ratio between the magnetic flux theading the horizon and the accretion rate~\cite{Tchekhovskoy:2011zx}, to the best of our knowledge the high-spin factor $f(\Omega_H)$ is always considered to be independent on the magnetosphere and the accretion disc models.

The findings presented here with the Kerr-MOG background support the idea that the high-spin factor $f(\Omega_H)$ is characteristic of the background BH metric, even though further research are needed in order to confirm this hypothesis in other alternative theories of gravity.
In this regards it would be relevant to extend the considerations and the results we derived here to theory-agnostic backgrounds like the Konoplya-Rezzolla-Zhidenko metric~\cite{PhysRevD.93.064015}, where a knowledge of the BZ high-spin factor $f(\Omega_H)$ is still missing. We plan to address this in a future project.

\section*{Acknowledgments}
I want to express my gratitude to the organizing commitee of the 17th Marcel Grossmann Meeting for the opportunity of presenting this research. 
It is my pleasure to thank the coauthors of the article that mainly inspired this talk, namely T. Harmark, M. Orselli and M. Rodriguez, as well as G. Grignani, O. Dias, J. Santos, R. Oliveri and A. Placidi for their fundamental contributions in previous works related to the same topic.
This work is supported by the ERC Advanced Grant ``JETSET:
Launching, propagation and emission of relativistic jets from binary mergers and across mass scales'' (Grant No.  884631).

\bibliographystyle{ws-procs961x669}
\bibliography{Bibliography}

\end{document}